# Investigation of W-SiC compositionally graded films as a divertor material

**Authors:** Zihan Lin[a,*], Carlos Monton[b], Stefan Bringuier[b], Gregory Sinclair[b], Guangming Cheng[c], Eduardo Marin[b], Zachary Bergstrom[b], Dmitry Rudakov[d], Žana Popović[e], Ulises Losada[f], Igor Bykov[b], Evan T. Ostrowski[a], Shota Abe[g], Nan Yao[c], Bruce E. Koel[a], and Tyler Abrams[b]

[a] Department of Chemical and Biological Engineering, Princeton University, Princeton, NJ 08544, USA
[b] General Atomics, San Diego, CA 92186-5608, USA
[c] Princeton Materials Institute, Princeton University, Princeton, NJ 08544, USA
[d] University of California San Diego, San Diego, CA 92093, USA
[e] Oak Ridge Associated Universities, Oak Ridge, TN 37830, USA
[f] Auburn University, Auburn, AL 36849, USA
[g] Princeton Plasma Physics Laboratory, Princeton, NJ 08543, USA
*corresponding author email: zhl030613@gmail.com

## Abstract

W-SiC composite material is a promising plasma-facing material candidate alternative to pure W due to the low neutron activation, low impurity radiation, and low tritium diffusivity of SiC while leveraging the high erosion resistance of the W armor. Additionally, W and SiC have high thermomechanical compatibility given their similar thermal expansion rates. The present study addresses the synthesis and performance of compositionally graded W-SiC films fabricated by pulsed-DC magnetron sputtering. Compositional gradients were characterized using transmission electron microscopy (TEM) and energy-dispersive X-ray spectroscopy (EDS), and crystallographic information was obtained using electron diffraction and X-ray diffraction (XRD). Samples were exposed to L-mode deuterium plasma discharges in the DIII-D tokamak using the Divertor Material Evaluation System (DiMES). Post-mortem characterizations were performed using scanning electron microscopy (SEM) and XRD. Electron diffraction and XRD showed that the compositionally graded W-SiC films were composed of polycrystalline W and amorphous SiC with amorphous W+SiC interlayers. No macroscopic delamination or microstructural changes were observed under mild exposure conditions. This study serves as a preliminary examination of W-SiC compositionally graded composites as a potential candidate divertor material in future tokamak devices.

# 1. Introduction

During transient and steady-state plasma discharges in a tokamak fusion device, divertor plasma-facing components (PFCs) must tolerate intense particle and heat fluxes. Tungsten (W) is a promising choice for the PFC in future fusion reactors due to its high melting point and thermal conductivity, excellent physical and chemical sputtering resistance, and strong creep resistance to withstand the damages imposed by the harsh divertor environment [1,2]. However, the current understanding is that the advantageous properties of W can undergo significant degradation due to neutron irradiation, limiting its performance in D-T fusion devices. High-energy neutron bombardment can introduce lattice defects and void formation [1]. Such structural deteriorations are accompanied by diminishing thermal and electrical conductivities and an increase in the ductile-to-brittle transition temperature (DBTT) [3], shortening the lifetime of the plasma-facing material. Additionally, transmutation processes leading to Re-rich phases have been observed in neutron irradiation experiments on W, which can lead to further embrittlement [1,3].

For several years, SiC has attracted considerable attention as a potential structural material in fusion reactors. SiC/SiC composites may enable high operating temperatures (~1,000 °C) and have low activation under high neutron flux [4,5]. Vapor-deposited β-SiC has also shown lower hydrogenic diffusivity relative to tungsten, which motivated the evaluation of SiC as a tritium diffusion barrier in PFCs [6–8]. The limitation is that low-Z materials like SiC are more vulnerable to sputtering compared to W as a plasma-facing material [9]. Consequently, there are safety concerns regarding high levels of tritium retention via codeposition with sputtered Si and C atoms [10,11]. Leveraging the high physical and chemical sputtering resistance of W and the neutronic properties and low tritium diffusivity of SiC, a W and SiC hybrid system may be considered as a potential divertor material candidate where W serves as the armor material and SiC as the neutron-resistant substrate and tritium diffusion barrier. W and SiC have several stable binary compounds at extreme temperatures (~1,800 °C), some with favorable thermal and mechanical properties [12–14]. Efforts to evaluate the feasibility of W-SiC composites have already been undertaken. Wright *et al.* [8] investigated SiC as a hydrogen permeation barrier in a W-SiC-W multi-layer construct fabricated via RF magnetron sputtering. The SiC permeation barrier effectively reduced D diffusing through the bulk W. Mohrez *et al.* [15] synthesized various W-SiC/SiC dual layers by the applied NITE process. The authors demonstrated that these dual layers survived divertor-relevant heat flux (~10 MW m$^{-2}$) in the Large Helical Device without significant visible damage.

The objective of this paper is to address material roughening and thermomechanical strains at the interface between W and SiC bulk layers in a W and SiC hybrid system, which can lead to macroscopic delamination. In the same study of W-SiC dual layer by Mohrez *et al.* [15], the W-SiC/SiC substrate layer was susceptible to crack propagation depending on the structure of and interactions with the W armor layer. Designing a compatible W-SiC binary system has also been initiated by PFC research at the DIII-D National Facility. Previously, a batch of W-SiC bilayer composites fabricated by pulsed-DC magnetron sputtering was exposed to L-mode plasma discharges using the Divertor Materials Evaluation System (DiMES). Due to the mechanical stress at the W-SiC bilayer interface, macroscopic delamination was observed. One solution was proposed by DelaCruz *et al.* [16], where the authors used a TiN diffusion barrier between W and SiC to suppress interfacial reactions and prevent mechanical failures such as delamination, but the mechanical behavior and the hydrogenic diffusivity of TiN in the context of fusion are not well understood.

The advantages of a W-SiC binary system and the challenges at the W-SiC interfacial regions warrant new designs for a W-SiC composite material. In this article, we report on the synthesis, characterization, and preliminary performance testing of compositionally graded W-SiC films in the lower divertor of the DIII-D tokamak [17,18]. The goal of the compositional gradient is to provide a more gradual transition in the composite material from a SiC-rich substrate region to a W-rich armor compared to a bilayer, thereby reducing interfacial stress. Two types of compositionally graded W-SiC films were fabricated using pulsed-DC magnetron sputtering, a technique that minimizes sample inhomogeneity and defects due to the charge accumulation on the target surface and the consequent arcing during deposition [19,20]. Cross-sectional images of compositionally graded W-SiC films were obtained via transmission electron microscopy (TEM). The samples were characterized by scanning electron microscopy (SEM), energy-dispersive X-ray spectroscopy (EDS), and X-ray diffraction (XRD) before and after L-mode deuterium plasma exposure. The primary purpose of this work is to serve as a foundation for developing functionally graded W-SiC composites as a candidate plasma-facing material.

## 2. Methods

*2.1. Fabrication and characterization of W-SiC compositionally graded films*

W-SiC compositionally graded thin films were deposited on 1.5 cm x 1.5 cm Si (100) wafers and crystalline cubic SiC-coated graphite substrates [9] via pulsed-DC magnetron sputtering. The working distance between the W (Plasmaterials, 99.95%) and SiC targets (Plasmaterials, 99.5%) and the sample holder was 11.5 cm. The substrates were mounted onto a rotating sample holder to limit lateral inhomogeneities in the coatings across all samples. Ar (99.999% purity) was used as the sputtering gas with a flow rate of 20 sccm. The chamber was pumped to $1.0 \times 10^{-7}$ Torr before introducing Ar. The deposition pressure during fabrication was kept constant at approximately 3.7 mTorr. Two different compositionally graded films (CGFs), hereafter known as CGF-1 and CGF-2, were fabricated according to the power profiles shown in Fig. 1, each targeting ∼ 4-5 µm thick films. The deposition rate at 100 W for W and SiC was around 0.5 µm $hr^{-1}$. TEM samples of the fabricated films were prepared with a focused ion beam in a Helios NanoLab G3 UC DualBeam instrument. Cross-sectional structural and elemental analyses were performed on the films deposited on Si wafers by TEM and EDS on a Titan Cubed Themis 300 double Cs-corrected Scanning/-Transmission Electron Microscope (S/TEM), using an accelerating voltage of 300 kV. Crystallographic information for the films deposited on Si wafers was obtained by electron diffraction on the same instrument.

TEM cross-sections of CGF-1 and CGF-2 deposited on Si (100) wafers in the high-angle annular dark-field (HAADF) mode are shown in Fig. 2. For each layer shown in Fig. 2a and 2b, the thickness is proportional to the deposition time and a constant deposition rate, which is proportional to the applied power. The strong mass thickness contrast and the layer thicknesses of the pure vs. codeposition regions are consistent with the fabrication conditions. EDS peaks of W (W Lα) at 8.398 keV, Si (Si Kα) at 1.740 keV, and C (C Kα) at 0.277 keV were used to verify the elemental gradient. The EDS peak for O (O Kα) at 0.525 keV was used to identify locations of potential impurities in the films during fabrication. The EDS maps showing the distribution of the different film layers are given in Fig. 3. CGF-1 consists of an approximately 1.5 µm SiC layer, followed by a 0.5 µm codeposition layer and a 1.5 µm W layer. The W gradient in both CGF-1 and CGF-2 shows the intended decreasing trend as a function of depth from the surface.

The elemental distribution of the pure SiC layer in CGF-1 determined from EDS is 43.5% Si, 32.0% C, 24.4% O, and 0.1% W. The gradient in CGF-2 is finer since the fabrication involved more and shorter pulse intervals. The pure SiC layer is approximately 0.2 µm, followed by a 3 µm codeposition layer and a 1.2 µm W layer.

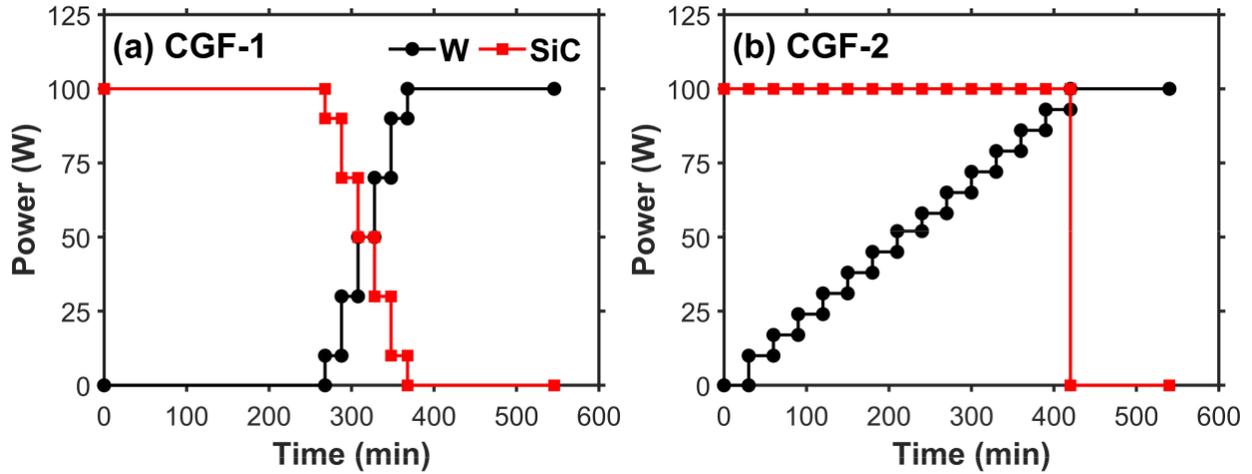

**Figure 1:** Applied power profile for sample fabrication via pulsed-DC sputtering deposition. The deposition was performed at constant pressure and the deposition rate can be assumed to be proportional to the applied power.

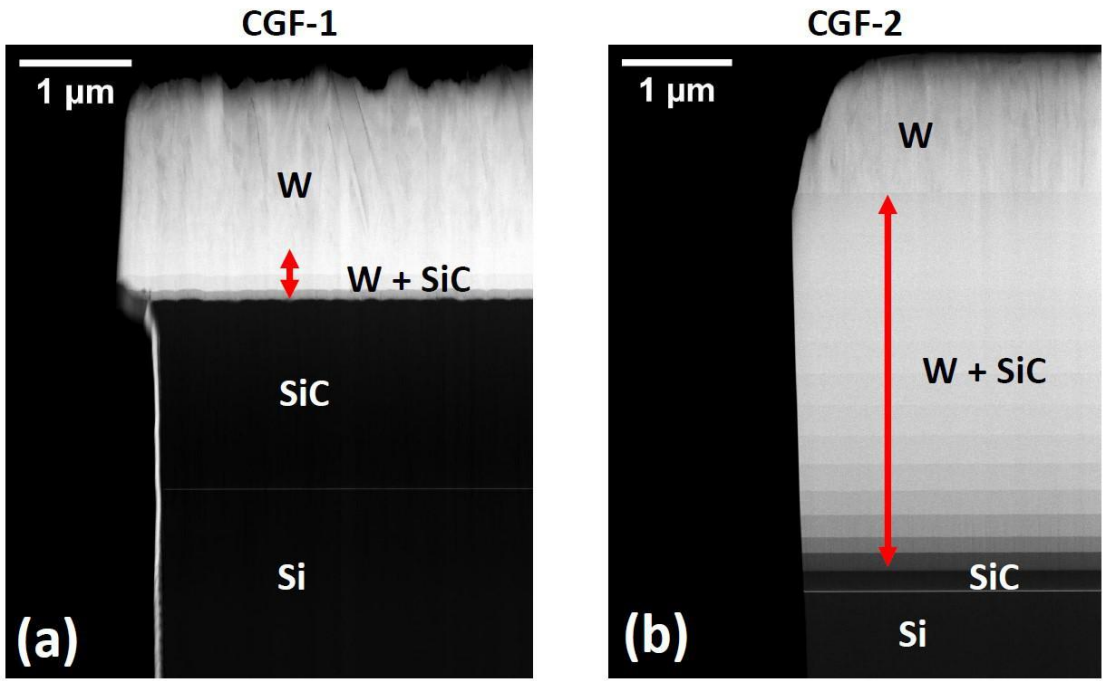

**Figure 2:** High-angle annular dark-field transmission microscopy images of (a) CGF-1 and (b) CGF-2.

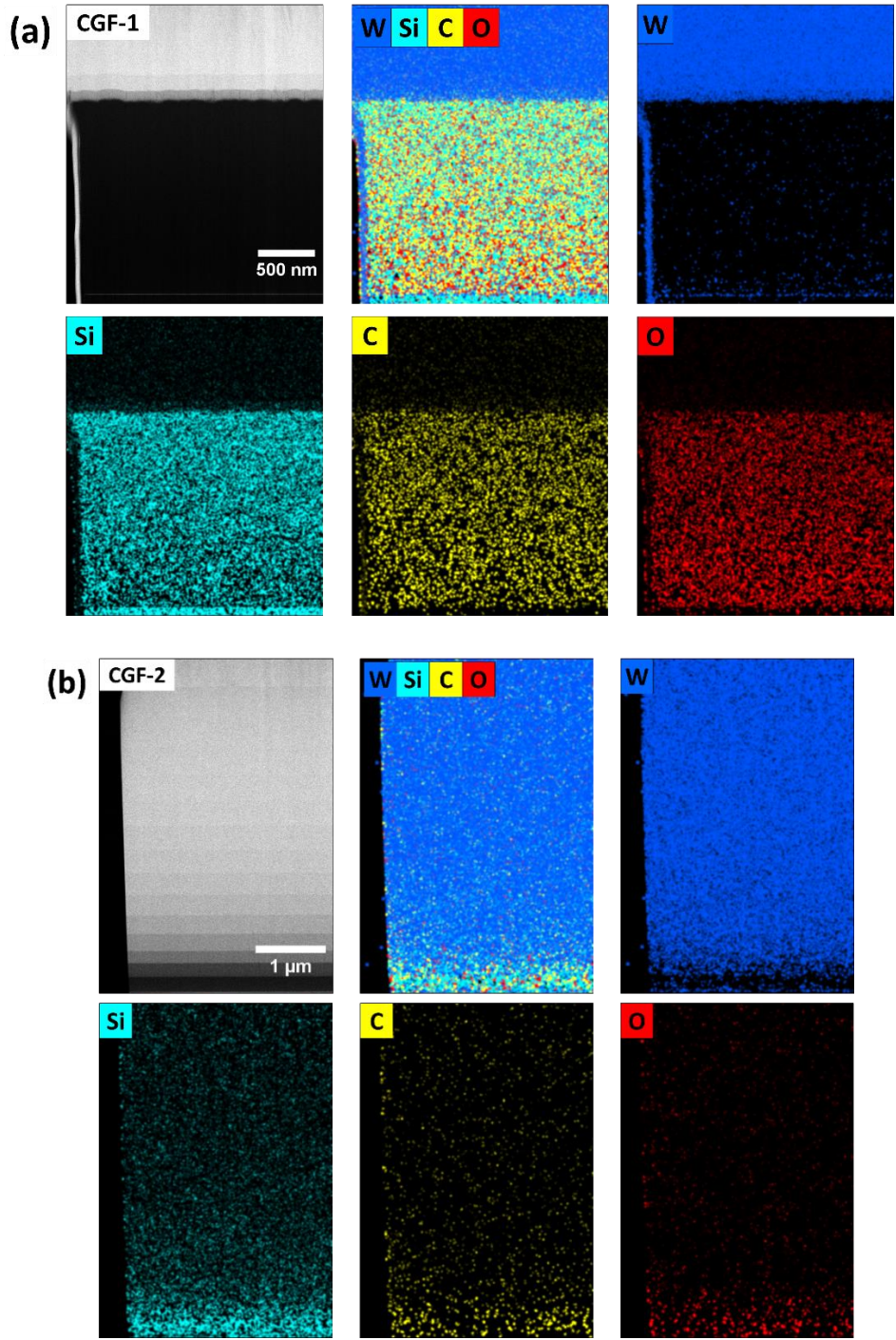

**Figure 3:** HAADF TEM images of the film cross-sections and EDS elemental maps of W, Si, C, and O for (a) CGF-1 and (b) CGF-2. The accelerating voltage was 300 kV.

*2.2. L-mode plasma exposure using DiMES and post-mortem characterization*

Compositionally graded films of W-SiC were deposited on crystalline cubic SiC-coated graphite substrates and exposed to L-mode deuterium plasma in the lower divertor of DIII-D using the Divertor Material Evaluation System (DiMES) [21]. DiMES allows for the insertion of multiple 6 mm wide circular sample surfaces flush with the divertor tile for studying plasma-materials interactions (Fig. 4a). The positions of the samples on the DiMES cap are shown in Fig. 4b. CGF-1A and CGF-1B refer to CGF-1 at two different radial positions. The nomenclature is identical for CGF-2. Divertor electron density and electron temperature were obtained using Divertor Thomson scattering (DTS), located 8 mm above the DiMES surface. The samples were exposed to 13 consecutive plasma discharges (shots 191404-191416) with the outer strike-point (OSP) on the divertor shelf, with an Ohmic heating power of ~ 1 MW and a plasma current of ~ 1 MA. The resulting plasma conditions are discussed in Sec. 3.2. In total, the plasma exposure time was ~ 42 s, with an average cumulative $D^+$ ion fluence of $6.3 \times 10^{19}$ ions cm$^{-2}$. Gross erosion rate of W was tracked by the spectral intensity of the W-I (400.09 nm) line using a multichordal divertor spectrometer (MDS) [22].

SEM (Quanta FEG-250 Scanning Electron Microscope) was used to examine the surface morphology of the films deposited on the DiMES substrates. Post-exposure samples were imaged on the DiMES probe, maintaining the orientation with respect to the toroidal magnetic field and radial directions during the plasma exposure. Surface elemental analysis was conducted after plasma exposure via EDS on the same instrument. XRD (Bruker D8 Discover X-Ray Diffractometer) was used to acquire the crystallographic information for the films deposited on the DiMES substrates using a Cu Kα radiation source (λ =1.5406 Å) over a scan range of 20–100° (2θ) at a scan rate of 1.5° min$^{-1}$ and a step size of 0.02°. SEM and XRD were both performed on samples before and after exposure to examine structural changes.

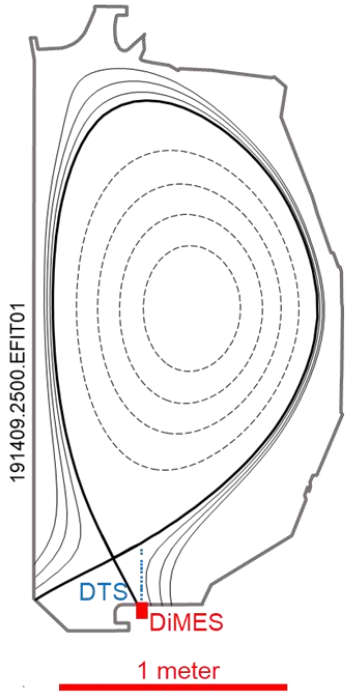
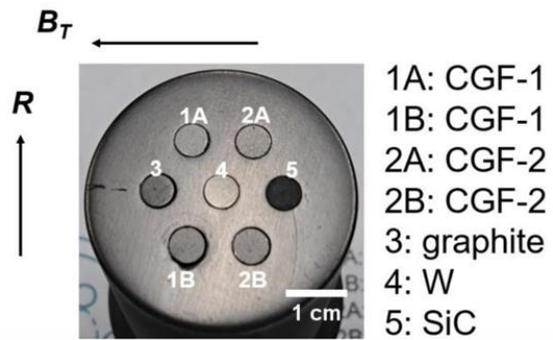

(a) DIII-D cross section
(b) Sample configuration

1A: CGF-1
1B: CGF-1
2A: CGF-2
2B: CGF-2
3: graphite
4: W
5: SiC

(c) Fabrication conditions

| Sample | Condition | Substrate |
| --- | --- | --- |
| CGFs | Varied power for 9 hours | Cubic SiC-coated DiMES sample |
| SiC | 100 W for 7 hours | Graphite DiMES sample |

**Figure 4:** (a) DIII-D poloidal cross section with typical magnetic equilibrium reconstruction overlaid. (b) Sample configuration on the DiMES apparatus before L-mode plasma exposure. (c) Processing conditions for samples that were fabricated by pulsed-DC magnetron sputtering for the DiMES experiment. Power profiles for the CGFs are given in Fig. 1.

# 3. Results

## 3.1. Electron diffraction and XRD characterization

Crystallographic information on CGFs deposited on Si and DiMES substrates was obtained via electron diffraction and X-ray diffraction, respectively. Fig. 5 shows the electron diffraction pattern of the cross-sections of CGF-1 and CGF-2 deposited on Si. The patterns indicate that the top W layer of the films is polycrystalline while the SiC and the codeposition layers (W + SiC) are amorphous.

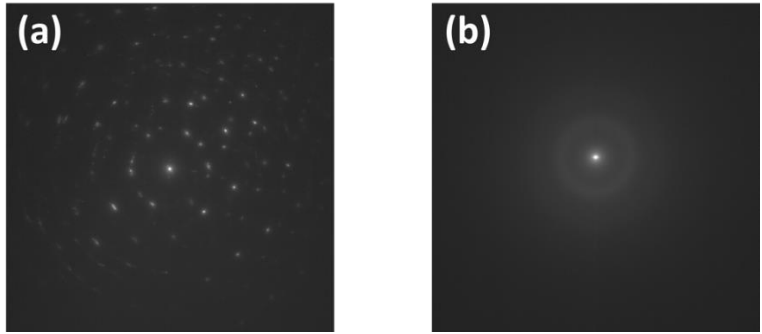

**Figure 5:** Electron diffraction pattern of (a) the top W layer in CGF-1 and (b) the W-SiC codeposition and SiC layers in CGF-2.

XRD patterns of films deposited on the SiC DiMES substrates before and after plasma exposure are shown in Fig. 6. The polycrystalline W electron diffraction patterns are assigned to BCC-W peaks at 40.28° (110), 58.20° (200), 73.20° (211), and 86.97° (220). The XRD patterns also inform the presence of crystalline cubic SiC, with peaks at 35.52° (111), 44.40° (200), 64.70° (220), 77.72° (311), and 98.37° (400). Since electron diffraction eliminated the deposited SiC as a source of crystalline peaks, we concluded that the XRD peaks assigned to cubic SiC belong to DiMES substrates that were previously coated with SiC [9]. The source of the broad feature from 20-40° was attributed to amorphous $SiO_2$ due to the native oxide layer on Si, as shown by EDS. We assigned the additional small peak at 26.62° to $WO_3$.

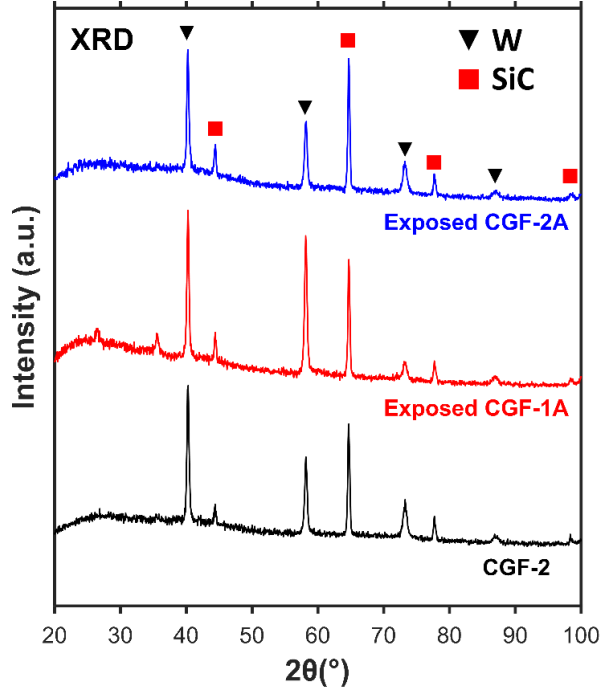

**Figure 6:** X-ray diffraction patterns of CGF-1 after exposure and CGF-2 before and after exposure.

*3.2. Divertor plasma conditions and post-exposure surface characterization*

The gross erosion rate of W from the CGFs was evaluated using the ionization per photon, or the S/XB method [22], a technique that involves converting calibrated absolute spectroscopic intensities into removal rates of atoms from the material surface. Since the direct exposure to the outer strike point was ~3.2 s per shot and the thickness of the top W layer in each sample is greater than 1 µm, we expect W to be the only product of erosion from the CGFs under the experimental conditions.

The spectral intensity of the W-I line (400.9 nm) was monitored by MDS for shots numbered 1, 2, 6-8, 12, and 13. We tracked Si-II (635.9 nm) and C-II (514 nm) emission lines for the other discharges in case the top W layer undergoes macroscopic delamination and subsequently reveals the W-SiC mixed interlayer and the SiC layer. DTS measured $T_e \sim$ 2-5 eV, with minor spikes up to 50-70 eV in shots 2-4, which is a low temperature regime. $D^+$ flux ($\Gamma_{D^+}$) during the L-mode plasma exposure was estimated by $\Gamma_{D^+} = 0.5(2T_e/m_i)^{1/2}\sin\theta$, where $m_i$ is the mass of $D^+$ ions, $\theta$ is the magnetic incidence angle, and $T_e = T_i$ was assumed [9]. The W erosion rate and $D^+$ ion flux for each discharge are shown in Fig. 7. The four CGFs experienced similar $D^+$ ion flux across different shots. A spike in W erosion was observed for shot 2, which is likely associated with the spike in $T_e$. Upon visual inspection of the exposed W-SiC CGFs, no macroscopic delamination or cracks were observed. Therefore, we did not perform TEM imaging of the cross-sections after plasma exposure. Fig. 8 shows the SEM images of the samples indicated in Fig. 4, which reveal similar surface morphologies before and after exposure. EDS surface elemental composition analysis of the samples is shown in Fig. 9 and Table 1, indicating the presence of carbon contamination from the graphite plasma-facing components native to DIII-D.

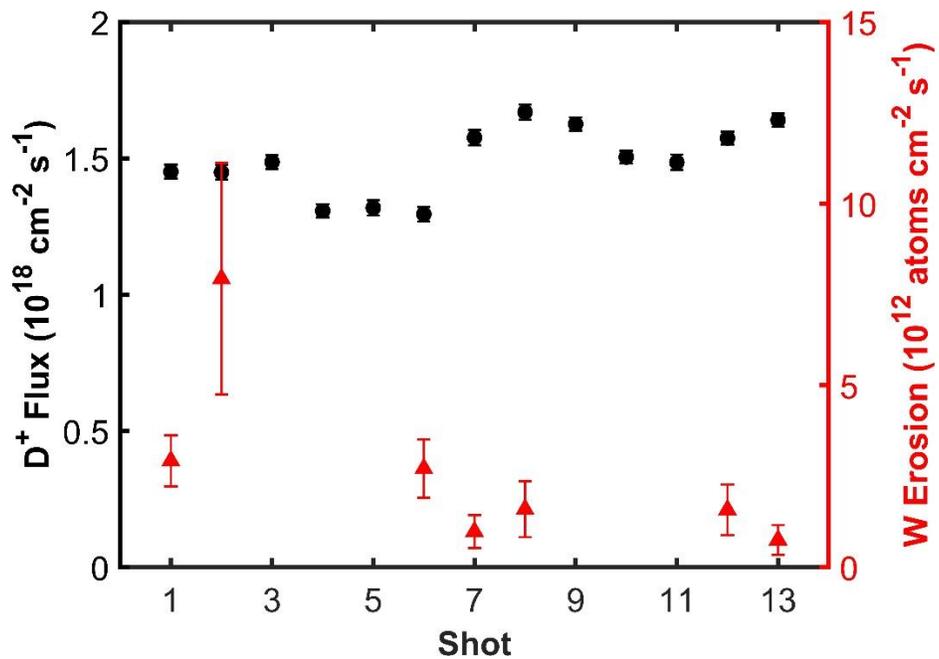

**Figure 7:** Divertor D$^+$ ion flux and W erosion flux from W-SiC CGFs for each shot. Erosion flux values were calculated using the ionization per photon method and averaged over the four W-SiC CGFs. The electron densities and electron temperatures were obtained from Divertor Thomson scattering located 8 mm above the DiMES surface.

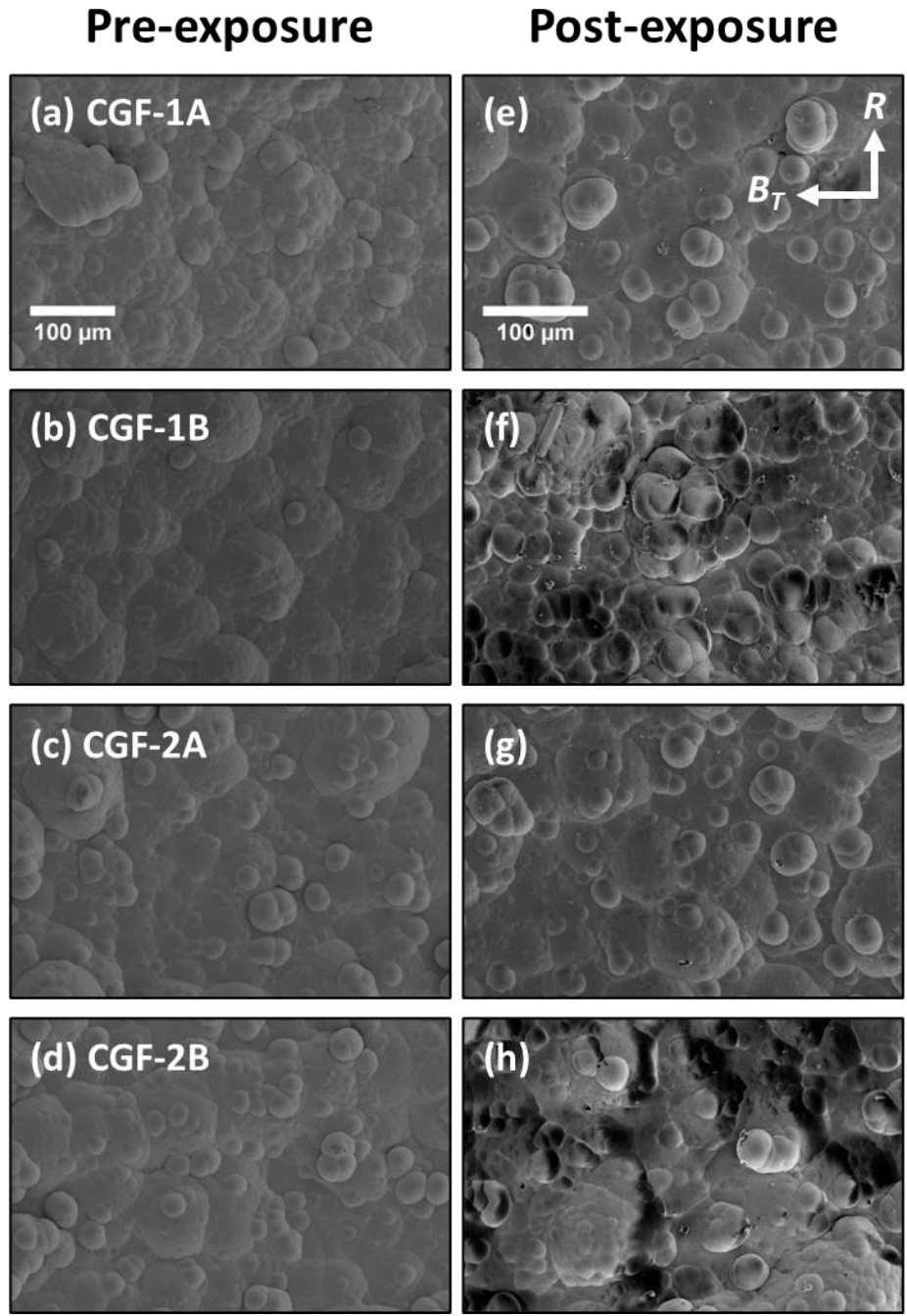

**Figure 8:** SEM images of the DiMES sample surface for (a) CGF-1A, (b) CGF-1B, (c) CGF-2A, (d) CGF-2B before and (e)-(h) after plasma exposure, respectively. Radial and toroidal magnetic field directions are shown. Images in each column are shown at the same scale.

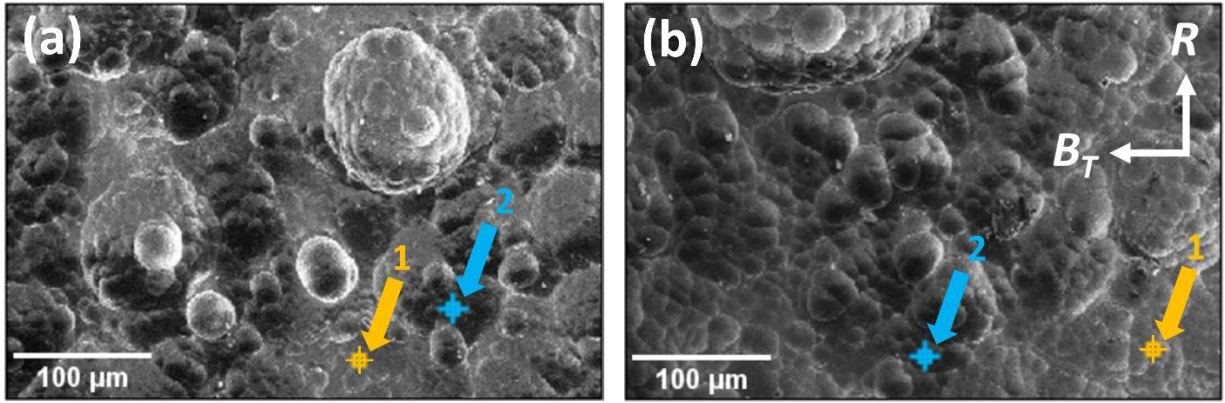

**Figure 9:** SEM images of the deposition/re-deposition spots on (a) CGF-1A and (b) CGF-1B. The accelerating voltage was 15 kV.

**Table 1:** Chemical composition of surface contamination from EDS.*

| Sample | | CGF-1A | | CGF-1B | |
|---|---|---|---|---|---|
| Measured spot | | point 1 | point 2 | point 1 | point 2 |
| Atomic percentage (%) | W | 74.9 | 21.1 | 72.5 | 40.0 |
| | C | 21.1 | 68.8 | 22.6 | 50.7 |
| | O | 3.9 | 10.1 | 4.9 | 9.3 |

*Points 1 and 2 for each sample are identified in Fig. 9.

## 4. Discussion

*4.1. Analysis of elemental composition and phases in the W-SiC films*

Cross-sectional TEM imaging and EDS elemental composition analysis show the intended compositional gradient in both CGF-1 and CGF-2. We observed some faint Si EDS signals in the pure W region in CGF-2, which was attributed to the difficulty of deconvolution of the Si Kα (1.740 keV) and W Mα (1.775 keV) transitions in the EDS energy spectra. Due to the proximity of these two X-ray emission lines, some W signals may be mistaken as Si signals. For both CGF-1 and CGF-2, the distribution for Si matches that of C as expected. Analysis of the oxygen elemental map shows that oxygen impurities are mostly accumulated in the SiC-rich region and heavily near the interface between Si and SiC. This is indicative of the native oxide layer at the Si surface before the films were deposited. EDS shows a higher concentration of Si compared to C (43.5% Si and 32.0% C) in the SiC layer in CGF-1. We attributed this elemental ratio to the presence of a mixture of SiC and $SiO_2$ in the layer. Within the interaction volume of the 300-kV electron beam, contributions of Si signals from sources other than the SiC, nominally from the Si wafer substrate and W signals from the codeposition layer above, are possible.

A comparison between CGF-1 and CGF-2 reveals that the interfacial regions between the pure SiC and the W + SiC codeposition region show different roughness depending on the thickness of the SiC layer. In CGF-1, where the pure SiC layer is thicker, the subsequent codeposition layer shows observable roughness in the HAADF TEM image (Fig. 2a). This is expected since the thickness of the SiC layer greatly exceeds the roughness of the Si substrate surface [23]. The thinner codeposition layers, therefore, are more conformal with the morphology of the SiC layer. In CGF-2, the thin individual layers conform to the morphology of the smooth Si substrate surface and do not exhibit apparent roughness.

Both the electron diffraction and XRD patterns indicate that no crystalline W-Si-C binary or ternary phases were formed. Therefore, the as-deposited films are found to be a mixture of polycrystalline BCC W and amorphous SiC, with amorphous W-SiC interlayers. We note that the deposited W in the interlayer is amorphous. The deviation of W from the stable BCC structure has been found in another study, where the growth of insufficiently thick W films by magnetron sputtering can lead to amorphous films [24]. Since we did not impose elevated temperatures on the substrate during film synthesis, we do not expect interfacial reactions between W and SiC, and the SiC layers are expected to be amorphous as a high substrate temperature is required for the growth of crystalline SiC films [25–29]. As formation of amorphous SiC was not intended, we aim in future studies to perform this deposition process at elevated substrate temperatures to form crystalline SiC, which is more suitable as a structural material.

*4.2. Post-mortem analysis*

The divertor plasma conditions did not cause macroscopic delamination or cracks on the W-SiC CGFs. No pronounced difference was observed in the surface morphology before and after exposure. Some visible dark spots were observed on the sample surfaces, which were attributed to carbon redeposition since the DiMES head is made of graphite and most of the contaminants in the DIII-D tokamak come from the graphite PFCs [30]. Microstructural analysis via SEM images shows that the carbon deposition on the sample surfaces correlates with the toroidal magnetic field direction, concentrating largely on the right side of the raised features on the sample surface. Moreover, the contaminant coverage is higher on CGF-1B and CGF-2B, which resulted from a higher concentration of C impurities from the graphite plasma-facing materials in DIII-D near the more radially inward samples, while the more specific impurity deposition patterns on the sample surface depend strongly on the ion incident angle distribution in the sheath [31]. A closer examination of the dark regions on CGF-1A and CGF-1B was conducted using EDS (Fig. 9). The brighter points on both samples correspond to regions with less carbon deposition, showing high W content and minor O contamination. The darker points correspond to areas of high carbon deposition. Oxidation of surface W after plasma exposure likely occurred due to reactions with oxygen and water impurities during plasma exposure at elevated temperatures and also after the samples were removed from DiMES and subjected to the ambient lab atmosphere, forming $WO_2$ or $WO_3$. Comparison between XRD patterns before and after exposure indicates that the plasma conditions did not induce changes in the crystalline structure. However, the emergence of tungsten carbides and silicides may be expected under harsher thermal conditions in future experiments [13,16,32]. The different W-Si-C phases that may form in different layers of the films will be investigated in future H-mode experiments.

**5. Conclusion**

In this paper, we focused on the synthesis and characterization of compositionally graded W-SiC films as a plasma-facing material for fusion applications. Deposition by pulsed-DC magnetron sputtering on cubic crystalline SiC-coated graphite and Si (100) wafer substrates produced polycrystalline W and amorphous W+SiC codeposition and SiC layers. The transition from a SiC-rich region to a W-rich region was characterized by TEM and EDS. EDS elemental analysis showed the desired W-SiC compositional gradient. The amorphous nature of the SiC was explained by the ambient substrate temperature during deposition. Based on the electron diffraction patterns, W codeposited with SiC in the W+SiC interlayers did not form crystalline W phases, which may be explained by insufficient thickness. The formation of stable phases upon interfacial reactions in the interlayers will be investigated in the future. Additionally, the observable roughness of the deposited layers from cross-sectional TEM imaging depends on the layer thickness and the roughness of the preceding layer.

A preliminary L-mode experiment using the DiMES apparatus in DIII-D shows that the film is stable under low-power plasma conditions. Under these conditions, plasma exposure did not lead to re-crystallization, macroscopic delamination, or apparent changes in surface morphology. The preliminary L-mode plasma exposure shows promising performance of compositionally graded W-SiC films compared to W-SiC bilayer configurations and motivates subsequent H-mode experiments to examine the performance and interfacial behavior of W-SiC compositionally graded films under higher particle and heat fluxes. Future work will also aim to achieve greater control of the crystallinity of the SiC layer, identify stable phases upon interfacial reactions, and quantify hydrogenic retention properties.

**Acknowledgment**
This material is based upon work supported by the U.S. Department of Energy, Office of Science, Office of Fusion Energy Sciences, using the DIII-D National Fusion Facility, a DOE Office of Science user facility, under Award(s) DE-FC02-04ER54698. The work was supported in part by US DOE under the Science Undergraduate Laboratory Internship (SULI) program under DE-FC02-04ER54698 and ICF Target Fabrication contract number 89233119CNA000063. The authors also acknowledge the use of Princeton's Imaging and Analysis Center (IAC), which is partially supported by the Princeton Center for Complex Materials (PCCM), a National Science Foundation (NSF) Materials Research Science and Engineering Center (MRSEC; DMR-2011750).

**Declaration of Competing Interest**
The authors declare that they have no known competing financial interests or personal relationships that could have appeared to influence the work reported in this paper.

and opinions of authors expressed herein do not necessarily state or reflect those of the United States Government or any agency thereof.